\documentstyle[12pt,a4wide]{article}

\newcommand{\vgl}[1]{eq.(\ref{#1})}

\newcommand{\cM}{{\cal M}}

\newcommand{\Ff}{\frac{1}{2\pi}}

\newsavebox{\uuunit}
\sbox{\uuunit}
    {\setlength{\unitlength}{0.825em}
     \begin{picture}(0.6,0.7)
        \thinlines
        \put(0,0){\line(1,0){0.5}}
        \put(0.15,0){\line(0,1){0.7}}
        \put(0.35,0){\line(0,1){0.8}}
       \multiput(0.3,0.8)(-0.04,-0.02){12}{\rule{0.5pt}{0.5pt}}
     \end {picture}}

\newcommand{\half}{{\textstyle\frac{1}{2}}}

\begin{document}
\begin{titlepage}
\begin{flushright} Preprint KUL-TF-93/3\\
                   hepth@xxx/9302015\\
                   January 1993
\end{flushright}

\vfill
\begin{center}
{\large\bf Klein-Gordon-Langevin Quantum Geometry.}
\vskip 27.mm
{\bf F. Vanderseypen$^\natural$}\\
\vskip 1cm
Instituut voor Theoretische Fysica
        \\Katholieke Universiteit Leuven
        \\Celestijnenlaan 200D
        \\B-3001 Leuven, Belgium\\[0.3cm]
\end{center}
\vfill
\begin{center}
{\bf Abstract}
\end{center}
\begin{quote}
\small
We present a quantum geometric framework for stochastic quantisation in the
case of a free Klein-Gordon field on Euclidean space. In this approach the
noise is part of the background space, spacetime is fuzzy. We extend the
notion of sharp point limit and show how fuzzy spacetime and the
Klein-Gordon field gives the Euclidean space and the stochastically
quantised Klein-Gordon field respectively.
\vspace{2mm}
\vfill
\hrule width 3.cm  {\footnotesize
\noindent $^\natural$ Aspirant N.F.W.O., Belgium,
 Francois\%tf\%fys@cc3.kuleuven.ac.be}\\
\normalsize
\end{quote}
\end{titlepage}

\section{Prologue}

\begin{center}
\footnotesize
`Antonio Stradivarius
voyait bien le probl\`eme soumis\\
 \`a tous les luthiers de son
temps.\\ Il savait aussi qu'il n'arriverait pas \`a le r\'esoudre\\ en
modifiant frileusement telle ou telle partie du violon de l'\'epoque \\
mais en redessinant celui-ci, bref en en modifiant la structure. \\
C'est ce qui arriva, un peu par hassard, en 1679 ..'\\[0.3cm]
{\sf Jean Diwo} {\em Les Violons du Roi,} 1992.
\end{center}
\normalsize
Stochastic quantisation is one of the many quantisation methods that
is available for quantising a field. It has the advantage that it is
simple ; for a given action $S$ one first solves the Langevin equation
\begin{equation}
\frac{\partial\phi(x,\tau)
}{\partial \tau}=-\frac{\delta S}{\delta\phi(x,\tau)} + \eta(x,\tau)
\label{Langevineq}
\end{equation}
in terms of the Gaussian white-noise field $\eta$
and the Green's functions are then the $\tau\rightarrow 0$ limit of the
$\eta$-expectation values
\begin{equation}
\langle \phi(x_1)\ldots\phi(x_n)\rangle = \lim_{\tau\rightarrow 0}\langle
\phi(x_1,\tau)\ldots\phi(x_n,\tau)\rangle_\eta.
\end{equation}
One also usually assumes that the background space is Euclidean
 \cite{Damgaard}.

It is well known that the white-noise $\eta$ in the description of the
Brownian motion of a `particle' in a fluid represents the average force
of the fluid on this particle \cite{Ford}. What is the analog in the case
of \vgl{Langevineq} ? If the field is the particle analog, it seems
natural to infer that the background space is fluid analog. Stated in
another way, can one make a connection between $\eta$ and the geometry
of the background space, has $\eta$ molecule-like properties ?
There is also the `time' parameter $\tau$, what is its origin ?
In the same way that the Brownian motion and the fluid fluctuations are
not present on a large scale, the geometry producing the random
 fluctuation $\eta$ should behave on a large scale as a solid space.
Such a geometry would carry the Langevin noise $\eta$, and as such would
quantise any field living on this geometry through the Langevin equation.
In this paper we argue that, in the case of a Klein-Gordon field, it is
possible to define a geometrical structure with the properties mentioned
above.
Many of the ideas presented below have their origin in the beautiful and
well-elaborated work by Prugove\v{c}ki (see \cite{QG} and references
therein) on Quantum Geometry.
In the next section we introduce the necessary concepts, of which the
quantum noise field is the most important, and define the sharp point limit
of a proces on stochastic phase space. This limit is used in section
three to show, first, how it damps out the vielbein fluctuations of the
space to produce a Ricci flat space and, second, how the stochastic
quantised Klein-Gordon field emerges from the diffusing quantumnoise field.
We conclude with some (yet) unsolved problems.
\section{About Spaces and Stochasticity.}
In this section we introduce two basic objects, the stochastic phase space
(abreviated as SPS) and the quantum noise field (abreviated as QNF). The
SPS is, loosely speaking, a set of points with fuzzy character.
This fuzzyness wil enable us to define the QNF, which will produce the
quantisation of the Klein-Gordon field (abreviated as KG field) and a
stochastic metric in the next section.

In the first part of this section we describe the spaces, in the second
stochasticity.
\subsection{Spaces}
{\sf The spacemanifold.} Assume that $\cM$ is an Euclidean manifold of
dimension $D$ and $q\in\cM$, in a certain coordinate system $q=(q^\alpha
)_{\alpha =1\ldots D}$. This is a classical or deterministic manifold in
the sense that to each point we can associate a delta distribution
describing the rigidness of this point. So, let us define a deterministic
point as a two-component object \begin{equation}
Q=(q,\chi^d_q(q')),
\end{equation}
where $\chi^d_q(q')=\delta (q-q')$ is called the confidence function and
represents the `probability' of the point $q\in\cM$. By `probability' I
mean that the delta function reflects the fact that the point has no
extension, it is fixed or sharp defined.\\
\\
{\sf The stochastic spacemanifold.} Suppose now that $\chi_q$ is a complex
(random) function which in some limit, in a sense to be made
precise
later, collapses to $\chi^d_q$. The collapsing procedure will be
called the classical
limit or sharp point limit. In this way we extend the spacemanifold
$\cM$ to a stochastic spacemanifold $\cM_s$ with points $Q=(q,\chi_q)$.\\
\\
{\sf Phasespace.} By $\Gamma_d(\cM)$ we denote the classical phasespace
associated to $\cM$. In a certain coordinate system $(q,p)=(q^\alpha
,p^\beta )\in \Gamma _d(\cM),\;\;\,\alpha ,\beta =1,\ldots,D.$\\
\\
{\sf Stochastic Phasespace.} In a similar way that we attached to each
$q\in \cM$ a confidence function, we attach to each $p$ of phasespace a
confidence function $\chi_p$.The elements
$\Pi $ of the stochastic phasespace are then constructed by a Cartesian
product
\begin{eqnarray}
\Pi&=& Q\times P\nonumber\\
&=& (q,\chi_q)\times(p,\chi_p)\nonumber\\
&=& ((q,p),\chi_q\chi_p)\nonumber\\
&=& (\pi,\chi_\pi).\nonumber\\
\label{SPSpunt}
\end{eqnarray}
In the next section we will give a concrete expression for $\chi_\pi$.
The confidence functions should be seen as an extra structure on a space in
the same way that a connection or a metric can be added on a manifold. It
should be noted that \vgl{SPSpunt} strongly suggests a description in terms
of fiber bundles\cite{QG}.

\subsection{Stochasticity.}
{\sf The $\eta$-proces.} At each point $\pi \in \Gamma_d(\cM)$ we attach a
multiplet
of random functions $\eta^a_\pi,\;\;a=1,\ldots,2D$ parametrized by $t\in R$
and $k\in R^D$. This random field on SPS is called the $\eta$-proces.
We make the following assumptions about the $\eta$-proces :
\begin{itemize}
\item the proces is independent of $q$ :
\begin{equation}
\eta^a_\pi=\eta^a_p.
\end{equation}
\item the proces is Gaussian, such that all the odd correlators are zero
while the even ones can be computed using the Wick-property and the fact
that
\begin{equation}
\langle\eta^a_p(k,t)\;\eta^b_{p'}(k',t')\rangle = \delta ^{ab}\delta
(p-p')\delta (k+k')\delta (t-t').
\end{equation}
\end{itemize}
{\sf The confidence functions.} The confidence functions we will use in
the rest of this work are
\begin{eqnarray}
\chi_q(k) &=& \left(\Ff\right)^{D/2}\exp-ikq\\
\chi_p^{\lambda ,a}(k,t) &=& \left(\frac{\lambda
^2}{\pi}\right)^{D/4}\;\eta^a_p(k,t)\;\exp-\half \lambda ^2k^2,
\end{eqnarray}
where $\lambda \in R$. The sharp point limit, denoted by $[\;\;]^\sharp$,
is now defined as
\begin{equation}
[\;\chi_p^{\lambda ,a}\;]^{\sharp}= \lim_{\lambda \rightarrow 0}\;\,
\langle
\chi^{\lambda, a}_{p}\rangle_\eta =0
\end{equation}
\begin{equation}
[\;\chi_{q}\;]^{\sharp} = \lim_{\lambda \rightarrow 0}\;\,\langle
\chi_q\rangle_\eta = \chi_q
\end{equation}
and we see that the confidence function of the $p$-submanifold collapses to
zero while
\begin{equation}
[\;\chi_q\;]^\sharp = \delta (q-q'),
\end{equation}
which means that we recover the deterministic manifold $\cM$.\\
\\
{\sf The quantumnoise field.} The confidence functions as used in the
context of Quantumgeometry (\`{a} la Prugove\v{c}ki) are nonstochastic and
play a fundamental role throughout the whole theory. I would like to show
that the use of random $\chi$-functions introduce new possibilities in
Quantumgeometry.
Define the
quantumnoise vector as
\begin{equation}
n^a(q,p,t)= \int dk\; \chi_q(k)\, \chi^{\lambda,a}_{p}(k,t),
\end{equation}
and note that it disappears
in the sharp point limit :
\begin{equation}
[\;n^a\;]^\sharp = 0.
\end{equation}
This last equation is of course a direct consequence of
\begin{equation}
\langle n^a(q,p,t)\rangle_\eta = 0,
\end{equation}
which seems to be a first indication that we could use $n^a$ in the sharp
point limit as a candidate for a Gaussian process. A short
computation shows that is indeed possible :
\begin{equation}
\langle n^a(q,p,t)\;n^b(q',p',t')\rangle_\eta= \left(\Ff\right)^{D}\delta
^{ab}\delta (t-t')\delta (p-p')\;\exp-\left[\frac{(q-q')^2}{4\lambda
^2}\right],
\end{equation}
such that in the sharp point limit
\begin{equation}
\left[\lambda ^{-D/2}\;n^a(q,p,t)\;n^b(q',p',t')\right]^\sharp =
\left(\Ff\right)^{D/2}\delta
^{ab}\delta (t-t')\delta (p-p')\delta (q-q').
\end{equation}
Which proves that the renormalized QNF behaves like a
Gaussian source in the limit $\lambda \rightarrow 0$.
\section{The Klein-Gordon System.}
In the first two parts of this section we describe how
 the QNF on SPS acts as a quantising stochastic source for the KG field on
a flat Euclidian manifold. The quantum field, however, is only the sharp
point limit of a diffusing field produced by the diffusion of the QNF.
In the last part we define several stochastic objects and show that in the
sharp point limit the Feynman propagator, the Euclidean metric and the
Ricci flatness is reproduced.
The idea that quantum field theory dictates geometry is not new,
see~\cite{Wettgeo} however for a nice approach.
\\
\\
{\sf Starting with diffussion on SPS ...}\\
\\
The classical diffusion equation on SPS is
\begin{equation}
\left[\frac{\partial }{\partial \tau}-\Delta \right]\psi(q,p,
\tau)=0,
\end{equation}
with $\Delta =\delta ^{\alpha \beta }\frac{\partial }{\partial q^\alpha}
\frac{\partial }{\partial q^\beta }$ the Laplace operator. Inserting the
QNF as a source on the right hand side gives
\begin{equation}
\left[\frac{\partial }{\partial \tau}-\Delta \right]\psi^a(q,p,\tau)=
n^a(q,p,\tau)
\label{diffnoise}
\end{equation}
and solving for $\psi^a$ results in
\begin{eqnarray}
\psi^a(q,p,t)&=& \int dk\;\chi_q(k)\;\Xi^{\lambda ,a}_p(k,t)\\
\Xi^{\lambda ,a}_p(k,t) &=& \int_0^t d\tau\; e^{-k^2(t-\tau)}\chi^{\lambda
,a}_p(k,\tau).\nonumber\\
\end{eqnarray}
A short inspectation shows that $\psi^a$ is nothing but the inverse
 Fourier
transform of the convolution of the Green function with $\hat{\chi}^{\lambda
,a}_p$. Let us put a mass term in \vgl{diffnoise} :
\begin{equation}
\left[\frac{\partial }{\partial \tau}-(\Delta-m^2) \right]\psi^a(q,p,\tau
)=n^a(q,p,\tau).
\label{diffnoisemass}
\end{equation}
This gives a modified $\Xi$ :
\begin{equation}
\Xi^{\lambda ,a}_{m,p}(k,t) = \int_0^t d\tau\;
e^{-(k^2+m^2)(t-\tau)}\chi^{\lambda,a}_p(k,\tau),
\end{equation}
as is easily checked.\\
\\
{\sf...passing along the quantum KG field...}\\
\\
{}From \vgl{diffnoisemass} it is now a small step to the Langevin
equation of the free KG field
\begin{equation}
\frac{\partial }{\partial \tau}\psi^a(q,p,t)=
(\Delta-m^2)\psi^a(q,p,t) +n^a(q,p,t)
\label{Langevin}
\end{equation}
with QNF as noise source. This stochastic differential equation produces a
certain `diffussion' field which, as proved above, will converge to the
quantised KG field in the sharp point limit since the stochastic noise has
Gaussian in this limit.
The solution for the stochastic differential equation is \begin{eqnarray}
\psi^a_{KG}(q,p,t)&=& \int dk\;\chi_q(k)\;\Xi^{\lambda
,a}_p(k,t)\\ \Xi^{\lambda ,a}_{\lambda ,p}(k,t) &=& \int_0^t d\tau\;
e^{(k^2+m^2)(t-\tau)}\chi^{\lambda,a}_p(k,\tau).\nonumber\\
\end{eqnarray}
 The linear combinations of
$\psi^a_{KG}$ with arbitrary normalised functions $c_a(p)$
\begin{equation}
\phi^\lambda _{KG}(q,t)=\int dp\; c_a\phi(p)\psi^a_{KG}(q,p,t)
\end{equation}
is also a solution of \vgl{Langevin}. Because the QNF behaves as a proper
Langevin/Gauss noise only in the limit $\lambda \rightarrow 0$ (and after
renormalisation), we see that the quantised KG field is the sharp point
limit of the renormalised $\phi^\lambda _{KG}$ :
\begin{equation}
\phi_{quant.}(q)=\lim_{t\rightarrow \infty}\;
\left[\lambda ^{-D/2}\phi^\lambda
_{KG}(q,t)\right]^\sharp.
\end{equation}
Note that all the renormalisation factors are infinite.\\

\newpage
{\sf...to end up up with geometry.}\\
\\
To make the connection between Langevin quantisation and geometry we define
a vielbein field on SPS as follows
\begin{equation}
e^A{}_\alpha (q,p,t)=\frac{\partial }{\partial q^\alpha }n^A(q,p,t),
\end{equation}
for $\alpha,A =1,\ldots,D.$
To define the inverse vielbein we should know
the precise expression of the vielbein, but we can proceed in another way.
Define the `stochastically inverse' vielbein as
\begin{equation}
e_A{}^\alpha =\delta _{AB}\,\delta ^{\alpha \beta }\,e^B{}_\beta ,
\end{equation}
and the stochastic metric in the usual way
\begin{equation}
g_{\alpha \beta }(q\mid p,p',t,t')=\left[\; e^A{}_\alpha\, \delta
_{AB}\,e^B{}_\beta\; \right]_{q=q'}
\label{smetric}
\end{equation}
Now, of course the objects defined above are ill defined since we
don't know the form of the $\eta$-process, but they make sense if we
average out the randomness, i.e. taking the sharp point limit :
\begin{equation}
\left[\;\lambda ^{-D} g_{\alpha \beta }\;\right]^\sharp\sim \delta
_{\alpha \beta },
\label{splmetric}
\end{equation}
\begin{equation}
\left[\;\lambda ^{-2D} g_{\alpha \beta }g^{\beta \rho
}\;\right]^\sharp\sim \delta _{\alpha \rho }.
\end{equation}
Here, just as before, we have to renormalise to get a sensible answer. All
this seems to indicate that the stochastic objects we defined before only
make sense in the deterministic limit, that quantisation is present because
of the fuzzyness but we have to average it out if we want a precise
expression. At this point the QNF is the origin of both the quantisation
and the metric fluctuation as defined above.
A last argument showing that the QNF behaves in a nice way as a geometric
object is the following\\
\\
{\bf Statement :}\\
{\sf The expectations of the stochastic
connection and the stochastic Ricci
scalar are zero, and a fortiori their sharp point limit :}
\begin{equation}
\left[\;\Gamma ^\alpha _{\beta \gamma }\;\right]^\sharp=0,
\label{connection}
\end{equation}
\begin{equation}
 \left[\;R\;\right]^\sharp=0.
\label{Ricci}
\end{equation}

We
note that although one expects these properties from~\vgl{splmetric} and
consistency requirements (remenber that we started with a flat Euclidean
space), the sharp point limit does not act transitively on a product such
that \vgl{connection} and \vgl{Ricci} are nontrivial results.

The stochastic connection and the stochastic Riemann tensor are, of course,
defined
in the usual way~\cite{QG}. The vanishing of the connection is then a
direct consequence of the Wick property and the fact that
\begin{equation}
\forall \alpha, \beta, \rho, A, B \,:\;\;\;\langle e^A{}_{\alpha,\beta
}\;e^B{}_\rho \rangle_\eta =0,
\end{equation}
which is easely verified,
while \vgl{Ricci} on the other hand is obtained after a long but
straightforward computation using the same tricks as for the connection.
\section{Conclusion.}
We have shown how the Gaussian noise in stochastic quantisation is related
to the geometry of the background space. The link between both is the
quantumnoise field, playing a double role : it is used as random source in
the Langevin equation and it produces a vielbein field on stochastic phase
space. In this way the `time' parameter, which seems to have only a minor
place throughout stochastic quantisation, has its origin in the fuzzy
points of the stochastic phase space.
There are, however, many open questions :
\begin{itemize}
\item can one proceed, even for this simple KG system, in a similar way if
the background space is curved ?
\item adding a potential to the KG system does not seem to give many
problems, but what about generalising to other theories ?
\item a derivation of the Langevin equation, based on `coarse graining' and
renormalisation, is still missing. Could a geometric principle really
produce the Langevin equation or the other way round ?
\item can one incorporate in some way the $t\rightarrow\infty$ limit into
the sharp point limit ?
\end{itemize}
\vspace{1.5cm}

I would like to thank P.Damgaard for helpful discussions, I'm also grateful
I had the opportunity to enjoy the facilities of CERN during this work.

\end{document}